# The effect of Hall drift on the Evolution of Magnetic Fields in the Crusts of Neutron Stars

Mikael Sahrling
Caltech, Mail Code 130-33, Pasadena, CA 91125
mikael@tapir.caltech.edu

## ABSTRACT

We consider the effect of Hall-drift on the evolution of magnetic fields in neutron star crusts using a finite difference code. Previously, most studies have focused on the situation where the ohmic term dominates the evolution. Here, we investigate the case where the ohmic term is small compared to the Hall term. We show that the effect on the magnetic field energy decay rate is a up to 30 % different compared to pure ohmic decay, depending on the initial conditions. In some cases the torroidal field act as an energy "reservoir" and cause a slower decay rate than the ohmic case. We also investigate the importance of high-order vacuum multipoles and find that for situations where the Hall term is a couple of orders of magnitude stronger than the ohmic term, hexapoles and up are a significant part of the magnetic field at the surface of the neutron star.

## 1. Introduction

Since the discovery of pulsars there has been much discussion of observational evidence for decay of their magnetic fields, as well as much theoretical work. As the reviews by Lamb (1991), Chanmugam (1992), and Phinney & Kulkarni (1994) indicate, there is at present no concensus on the question of whether or not magnetic fields in isolated neutron stars can decay significantly. The general view has been that the electrical conductivity of matter in the cores of neutron stars is so high that the characteristic decay time for fields generated by electrical currents in the core is greater than the age of the Universe. Many scenarios have been proposed to try to lower this decay time. For example Haensel, Urpin & Yakovlev (1990) have drawn attention to the possibility that magnetic fields in the core could decay rapidly by ambipolar diffusion. This process has been discussed from a microscopic point of view by Yakovlev & Shalybkov (1991a, b). However, Pethick (1992a) and Goldreich & Reisenegger (1992) have argued that, as a consequence of the relatively long times required for establishment of $\beta$-equilibrium, ambipolar diffusion can be hindered by the build-up of chemical potential gradients that reduce the counterflow of charged particles (electrons and protons) with respect to neutral ones (neutrons).

Another interesting possibility is the effect of Hall drift in the neutron star, see Jones (1988), Goldreich & Reisenegger (1992), Naito & Kojima (1994), Muslimov (1994). A neutron star is one of the few places where this term in the induction equation is relevant, due to the high magnetic field strength and high conductivity of those stars. From dimensional arguments Goldreich & Reisenegger (1992) speculated that turbulent cascades due to Halldrift can push the field to length scales that are several orders of magnitude smaller than originally. Locally then, the ohmic decay can be much faster than at first expected. However, the global decay is determined by the speed with which the field is transported to small scales. Most other studies have been done in the regime where the field strength is such that the ohmic term dominates the evolution, Naito & Kojima (1994), Muslimov (1994). They approached the problem by doing a multipole expansion of the field up to order $l = 10$. However, since the Hall term can be a couple of orders of magnitude stronger than the ohmic term one should include multipoles up to $l =$ a few $\times 100$. This is very cumbersome to do by multipole expansion and instead I have chosen to develop a finite difference code in 2 space- and 1 time dimension (axisymmetric case) to tackle the problem.

In this paper we examine cases where the field is a dipolar, poloidal field with two different radial dependencies initially. There are many ways to investigate the effect of Hall drift, and I have chosen to focus on aspects that are of astrophysical importance and leave many of the interesting theoretical questions aside for the moment. The paper is organised as follows: In section 2 the basic equations and boundary conditions are derived followed by section 3 where the numerical method is desribed. Section 4 describes the initial conditions used and investigates the magnetic energy evolution with and without Hall drift and the multipole evolution. The conclusions are finally presented in section 5.

## 2. Magnetic field evolution

The basic model I am considering is an axi-symmetric spherical shell of outer radius $R_2$ and inner radius $R_1$. Below $R_1$ there is a superconductor and outside the shell is vacuum. In the shell I assume the density and temperature are such the atomic nuclei are frozen into a lattice and the electrons form a completely relativistic, and degenerate gas. For simplicity let us consider the case where the conductivity and density in the shell are constant. I assume further that the magnetic field vanishes within the London penetration depth in the superconductor.

The fundamental equations we are solving are the magnetic induction equation, and Ampère's law

$$\frac{\partial \mathbf{B}}{\partial t} = -c\nabla \times \mathbf{E} \qquad ; \qquad \mathbf{j} = \frac{c}{4\pi}\nabla \times \mathbf{B} \tag{1}$$

with an electric field given by

$$\mathbf{E} = \frac{\mathbf{j}}{\sigma} + \left(\frac{\mathbf{j}}{en_c c}\right) \times \mathbf{B} \quad , \tag{2}$$

where $\sigma$ is the electric conductivity and $n_c$ is the electric charge density. The other constants have their usual meaning. For a derivation see e.g. Goldreich & Resisenegger (1992). In the crust the current is carried by the electrons and one can put $\mathbf{j}/(en_c) = -\mathbf{v}_e$, where $\mathbf{v}_e$ is the electron mean velocity. Using this fact and equation (2) for the electric field we get for the induction equation

$$\frac{\partial \mathbf{B}}{\partial t} = -c\nabla \times \left(\frac{\mathbf{j}}{\sigma}\right) + \nabla \times (\mathbf{v}_e \times \mathbf{B}) \quad . \tag{3}$$

The first term on the right hand side of equation (3) describes ohmic diffusion while the second term describes advection of the field by Hall drift. A useful parameter to distinguish between situations in which ohmic diffusion dominates the evolution of the magnetic field and those in which advection dominates is the magnetic Reynold's number $R_m$, see Jackson (1975). From a dimensional analysis of equation (3) we find that in our model

$$R_m = \frac{\sigma v_e B}{jc} = \frac{\sigma B}{en_c c} \ , \tag{4}$$

where $j, B$, and $v_e = j/(en_c)$ are scale factors appropriate to the largest magnetic structures. The advection term dominates the evolution of the magnetic field when $R_m > 1$. In this paper we will examine values of the magnetic Reynold's number ranging from $R_m = 5$ to $R_m = 100$. Previously, most studies have focused on situations where $R_m \leq 1$.

The magnetic field can be divided into a poloidal, $\mathbf{B}_{pol}$, and a torroidal component, $B_\varphi$. For the axisymmetric case that we are solving it is also advantageous to use the vector potential, $\mathbf{A} = A_\varphi \mathbf{e}_\varphi$, to get $\mathbf{B}_{pol}$. The magnetic field is then given by $\mathbf{B} = \nabla \times (A_\varphi \mathbf{e}_\varphi) + B_\varphi \mathbf{e}_\varphi$. This way of splitting the magnetic field is convenient because, together with the axi-symmetry, it automatically fulfills the $\nabla \cdot \mathbf{B} = 0$ condition from Maxwell's equations.

It is clear from looking at equation (3) that there are two time scales in the problem,

$$t_{ohm} = \frac{4\pi\sigma L^2}{c^2} \quad ; \quad t_{Hall} = \frac{4\pi e n_c L^2}{cB} \quad , \tag{5}$$

where $L$ is a length scale over which the magnetic field varies. The ratio between the two time scales equals the magnetic Reynold's number, $t_{ohm}/t_{Hall} = \sigma B/(en_c c) = R_m$.

To solve a particular problem boundary conditions are needed and in the next section we will discuss what they are for our purpose.

### 2.1. Boundary Conditions - General

From the magnetic induction equation we find that the boundary conditions to be imposed are the continuity of the tangential components of the electric field across boundaries, see for example Landau et al. (1984). In the geometry of our model the components $E_\varphi$ and $E_\theta$ must be continuous across $r = R_1, \ R_2$. We get using (2) :

$$\frac{j_\theta}{\sigma(\mathbf{r})} + \frac{j_\varphi B_r - j_r B_\varphi}{en_c(\mathbf{r})c} = E_\theta^{external} \tag{6}$$

$$\frac{j_\varphi}{\sigma(\mathbf{r})}+\frac{j_r B_\theta - j_\theta B_r}{en_c(\mathbf{r})c}=E_\varphi^{external} \tag{7}$$

where the super script "external" refers to either the field in the super conducting side at $r = R_1^-$ or the electric field in the vacuum at $r = R_2^+$.

From $\nabla \cdot \mathbf{B} = 0$ we have in our geometry that the radial component of $\mathbf{B}$, $B_r$ is continuous across the boundaries.

Finally, Ampère's law gives, in the absence of surface currents, that the tangential components of the magnetic field, $B_\theta$ and $B_\varphi$, also need to be continuous across the boundaries. In our model the inner boundary has a superconductor where surface currents can run and this condition only applies to the outer boundary.

### *2.1.1. Inner Boundary Conditions*

Assume now that the core is a superconductor (type I). The electric and magnetic field is initially zero and will therefore stay zero, which means that no magnetic flux can leak into the core beyond the London depth. The boundary conditions for the inner boundary are then:

$$B_r = 0 \tag{8}$$

$$\left.\left(\frac{j_\theta}{\sigma(\mathbf{r})}-\frac{j_r B_\varphi}{en_c(\mathbf{r})c}\right)\right|_{r=R_1} = 0 \tag{9}$$

$$\left.\left(\frac{j_\varphi}{\sigma(\mathbf{r})}+\frac{j_r B_\theta}{en_c(\mathbf{r})c}\right)\right|_{r=R_1} = 0 \tag{10}$$

When using the vector potential to describe the polar component of the magnetic field equation (10) simply means $\partial A_\varphi/\partial t|_{r=R_1} = 0$. From equations (8) and (10) and using the axisymmetry we find that $A_\varphi(t=0)|_{r=R_1} = 0$ and the inner boundary condition for the vector potential is then

$$A_\varphi(t)|_{r=R_1} = 0 \quad . \tag{11}$$

Equation (9) is the boundary condition for $B_\varphi$. Note that the boundary conditions mean that you have current sheets running along the border ($j_r$ does not have to be zero). This in turn implies that the boundary condition requiring $B_{\theta,\varphi} = 0$ is not valid at the inner boundary.

### *2.1.2. Outer Boundary Conditions*

We are assuming that there is vacuum outside the star, $r > R_2$. With no current sheets in the outer boundary the components of the magnetic field should be continuous across the outer boundary.

To further understand the outer boundary condition let us derive the field equations in the vacuum by investigating how the various terms in Maxwell's equations scale and use the magnetic field strength as a fundamental variabel, $B$. The electric field strength $E = |\mathbf{E}|$ is then $E \sim v/cB$ where $v$ is the speed given by the current in the crust, $v = j/(en_c)$. Putting these scalings into Maxwell's equations we find $c\nabla \times B = \partial E/\partial t \sim Ev/R \sim v^2 B/(cR) \Rightarrow \nabla \times B = (v/c)^2 B/R$. To second order in $v/c$ we then get $\nabla \times \mathrm{B} = 0$. This is due to the large conductivity of the neutron star crust. On the other hand, $\partial B/\partial t = Bv/R = -c\nabla \times E = -Bv/R$ which simply means that Faraday's law stays as it is. We have now for the field equations outside the star,

$$\frac{\partial \mathbf{B}}{\partial t} = -c\nabla \times \mathbf{E}\ , \quad \nabla \times \mathbf{B} = 0 \tag{12}$$

$$\nabla \cdot \mathbf{E} = 0\ , \quad \nabla \cdot \mathbf{B} = 0\ . \tag{13}$$

Combining $\nabla \cdot \mathbf{B} = 0$, $\nabla \times \mathbf{B} = 0$ we can define a scalar potential $\Phi_B$ such that

$$\mathbf{B} = \nabla \Phi_B \qquad ; \qquad \nabla^2 \Phi_B = 0. \tag{14}$$

This means $B_\varphi = 0$ outside because of the azimuthal symmetry. The solution for $\Phi_B$ is, using the boundary condition $\Phi_B \to 0$ as $r \to \infty$,

$$\Phi_B = \sum_{l=1}^{\infty} A_l r^{-(l+1)} P_l(\cos\theta) \tag{15}$$

where $P_l(\cos\theta)$ is a Legendre polynomial of order $l$. The boundary condition to be imposed are now (no current sheets at the outer surface)

$$B_r, B_\theta \text{ continuous at } r = R_2\ , \tag{16}$$

$$B_\varphi|_{r=R_2} = 0\ . \tag{17}$$

Note here that we are not interested in how the electrical field outside the star evolves and then we do not need the boundary conditions corresponding to equations (6), (7). The boundary condition for $A_\varphi$ can be found by combining the outer solution, $\Phi_B$, with the requirement that the poloidal components should be continuous across the boundary. We find,

$$\frac{\partial r A_\varphi}{\partial r} = -\sum_l \frac{1}{l+1}\left(\int_0^\pi A_\varphi \frac{\partial P_l}{\partial \theta} \sin\theta \mathrm{d}\theta\right)\frac{\partial P_l}{\partial \theta}\ , \tag{18}$$

With this boundary condition there is no need to solve for $\Phi_B$ outside.

## 3. Initial Conditions

For initial conditions I chose dipolar fields with two different radial functions. This has the advantage that we can investigate a weak and strong initial current. In the first case the lowest order ohmic eigenmode was chosen. This I then compared to a case where the field is initially buried inside the crust. Further, for each of these intial functions I chose five different Reynold's numbers ranging from 5 to 100 initially. The stronger the initial Reynold's number the stronger the Hall effect. In both cases the torroidal field was initially chosen to be zero.

### 3.1. Lowest Ohmic Eigenmode

A natural choice for initial conditions are ohmic eigenmodes. The temperature evolution of the neutron star crust suggest that in the first few years of a neutron star's life $R_m$ is small due to the initially low conductivity of the crust, Pethick (1992b). Therefore all high-order multipoles decay away quickly and only the lowest order eigenmode remains when impurity scattering starts to dominate the conductivity and $R_m$ is high. The lowest order eigenmode is also the one with smallest initial current and the subsequent drag of field lines into the torroidal component is expected to be weak. Using the vector potential we find,

$$A_\varphi(t=0) = (Aj_1(kr) + Bn_1(kr))\sin\theta \quad ; \quad B_\varphi(t=0) = 0, \tag{19}$$

where $j_1$, $n_1$ are the spherical Bessel and Hankel functions of order 1, respectively. The constants $A, B$, and $k$ are given by the boundary conditions and the choice of $R_m$.

### 3.2. Buried

In the past year there has been some attention given to initial field configurations that are buried within the crust. The subsequent evolution of the field due to ohmic diffusion creates higher and higher surface field which eventually dies off, Muslimov & Page (1995), Young & Chanmugam (1995), and Chanmugam & Sang (1989). The idea is that accretion buries the field inside the crust which can explain the low field in millisecond pulsars, which are thought to have been spun up by accretion, Phinney & Kulkarni (1994). Here I examine this scenario with Hall drift. Since the field variation is more rapid in this case, the initial current is stronger.

$$A_\varphi(t=0) = \frac{A}{r}\sin^2\left(\frac{\pi}{(R_2-R_1)}(r-R_1)\right)\sin\theta \quad ; \quad B_\varphi(t=0) = 0 \tag{20}$$

The boundary conditions are automatically satisfied and I only have to match $A$ to the chosen $R_m$.

## 4. Numerical method

I have chosen to solve equation (3) with boundary conditions given by equations (9), (11), (17), and (18) by a finite difference method using a grid with azimuthal symmetry and uniform in $r$-$\theta$, the poloidal variables, with $N_r$ gridpoints in the $r$-direction and $N_\theta$ gridpoints in $\theta$. The derivatives are calculated to second order in $\delta r = (R_2 - R_1)/(N_r - 1)$ and $\delta\theta = \pi/(N_\theta - 1)$. The time step, $\delta t$, is given by the Courant condition, $\delta t = \min_{j,k}(\delta r/v_{r,jk}, r\delta\theta/v_{\theta,jk})$, where $v_r$ and $v_\theta$ are the radial and angular components of the electron mean velocity at grid point $j,k$.

The method I have employed is operator splitting, where first the terms with radial derivatives in the variable considered are updated, followed by an update of the terms with angular derivatives. The calculation is organized so the vector potential $A_\varphi$ is updated at timestep $n$ and the torroidal field, $B_\varphi$ at timestep $n+1/2$. The accuracy in time is then of order $\delta t$, see Bowers & Wilson (1991).

To check the accuracy of the code I have performed several tests. First, turning of the Hall-drift operators and keeping only the ohmic part I found the relative error for each gridpoint to be less than 1 % after 5 ohmic time scales when using a timestep given by the condition, $\delta t = 4\pi\sigma\delta r^2/c^2$. To test the Hall-drift solver I simply turned off the ohmic part and made sure that the energy was conserved. After some ten Halldrift timescales the energy was conserved to a few 0.1%, see figure 1. The final test was done by using both ohmic and Hall operators and keeping track of the energy lost through ohmic decay and adding it to the total magnetic energy in the crust and outside in the vacuum.

$$E_{tot}(t) = \int_\infty \frac{|\mathbf{B}(t)|^2}{8\pi} dV + \int_0^t \int_V \frac{|\mathbf{j}(t')|^2}{\sigma} dV\,dt' \tag{21}$$

This sum has to be constant since the only source of energy loss is via ohmic decay. In the result section the accuracy of the various runs is given by

$$\epsilon = \sqrt{\frac{\left[\sum_{n=1}^{N} \left(E_{tot}(t_n) - E_{tot}(t=0)\right)^2\right]}{E_{tot}(t=0)^2 N}} \quad , \tag{22}$$

where $t_n$ is the time at timestep $n$ which is generally every 500 timesteps, and $N$ is the total number of such sampled timesteps.

## 5. Results

I assumed the charged particle density, $n_c$ and conductivity, $\sigma$ in the crust to be constant and equal to the values at the inner crust. This can be justified by the fact that the energy dissipation rate of the field is largely determined by the inner crust (Pethick & Sahrling 1995). Assuming impurity scattering dominates the conductivity and an impurity content of about 10 %, one gets $\sigma = 10^{27}$ s$^{-1}$ from Urpin & Yakovlev (1980) and assuming a proton fraction of about 5 % that of the total mass we end up with $n_c = 6 \times 10^{36}$ cm$^{-3}$. The inner radius was chosen to be $R_1 = 9$ km and the outer radius $R_2 = 10$ km. In all calculations I used $N_r = 100$ and $N_\theta = 100$.

The energy evolution in the crust with and without Hall drift is discussed in section 5.1. Section 5.2 discusses the vacuum multipole evolution.

## 5.1. Energy Evolution

In this section the magnetic energy evolution in the crust is discussed as a function of time. It is done with and without the Hall-term in equation (3) so a direct comparison can be made with pure ohmic decay.

### *5.1.1. Lowest Eigenmode Case*

In table 1 I have put the ratio of the total magnetic energy in the crust with Hall drift to the magnetic energy without Hall drift (only ohmic operator) as a function of time for a number of different initial values of $R_m$. The ohmic time is here given by solving the eigenvalue problem which gives $t_{ohm} = 5.33\ 10^{16}$ s. As one can see from the table the difference can be some ten percent for $R_m = 20$ to 100. The case with $R_m = 100$ was very time consuming and was stopped after only one ohmic time had passed. Figure 2 shows the run with $R_m = 100$ initially. It is interesting to see the small amount of energy pushed into the torroidal field, in particular when comparing to the buried case, figure 3, where the initial current $j_\varphi$ is greater.

### *5.1.2. Buried case*

In table 2 the ratio of the magnetic energy with Hall drift to the magnetic energy without Hall drift is tabulated as a function of time for a number of different initial values of $R_m$. The ohmic time in the table is here given by equation (5). With $L = 10^5/4$ cm, one gets $t_{ohm} = 8.74\ 10^{15}$ s. This is the ohmic decay time initially but since the initial condition is a sum over ohmic eigenmodes the ohmic evolution can not be described by a single timescale. The case with $R_m = 100$ was also here very time consuming and a very sharp feature developed at the inner boundary, which forced the calculation to be stopped after about one $t_{ohm}$. One interesting feature is the that the torroidal component act as an energy reservoir for the total energy. From figure 3 where the $R_m = 50$, 100 cases are shown with the poloidal and torroidal energy components displayed separately, one can see that after about an ohmic time the torroidal component pushes energy back into the poloidal component. Bear in mind that this is likely to depend strongly on the boundary and initial conditions used. The total effect on the energy is a few 10 % compared to the pure ohmic case.

### *5.1.3. Discussion*

One important state in the evolution of the magnetic field, starting from a purely poloidal field, is when the poloidal current, $j_{pol} \sim B_\varphi / l_\varphi$ is roughly equal in magnitude to the torroidal current, $j_\varphi \sim B_{pol}/l_{pol}$ , where $l_\varphi$ and $l_{pol}$ are the torroidal and poloidal length scales, respectively. In our case this is likely to occur when the torroidal field strength is less than the poloidal field strength, which can be a clue to why only a fraction of the total energy is pushed into the torroidal component.

To see this in more detail let us examine the evolution equation for the torroidal field. The Hall term consists of terms quadratic in $B_{pol}$ or $A_\varphi$. These terms act as a source for the torroidal field. Thus, if $A_\varphi$ varies with some period in some direction, the torroidal field created will vary with half that period and $l_\varphi < l_{pol}$. To match the torroidal and poloidal currents will thus require a field strength $B_\varphi = B_{pol} l_\varphi / l_{pol} \approx B_{pol}/2$. The corresponding torroidal energy is then a few tenths of the poloidal energy, which is the numerical result. It is then plausible that the creation of a poloidal current can be a limiting factor for the torroidal energy evolution. However, the above argument is of course limited in the sense that it assumes you get a strong poloidal current where the torroidal current is strong, something that does not have to happen everywhere in the crust. Also from the numerical results we know that the torroidal energy is strongly dependent on the initial conditions, something not accounted for with this argument.

## 5.2. Vacuum Multipole Evolution

The evolution of the vacuum multipoles are interesting since from observational arguments one expects the dipole field of recycled pulsars, i.e. pulsars that have been spun-up by accretion, to dominate near the neutron star surface, with possible addition of up to some 40 % of a few higher order ones, see Arons (1993) and the references therein. The initially buried case is the one in our model that represents the post accretion evolution. Note that in our model both the density and conductivity are constant and one should be careful about drawing too stringent conclusions regarding the effect on more realistic neutron star crust models.

In our model one can show from the symmetry of both initial conditions and the basic equations that only odd multipoles will appear as a function of time.

To show the importance of vaccum multipoles I use the surface field strength at the pole, $\theta = 0$. Here the field is purely radial and we have, $B_l = \partial \left( A_l r^{-(l+1)} \right) / \partial r = -A_l (l+1) R_2^{-(l+2)}$, see equation (15). Tables 3 and 4 shows the maximum of $|B_l(t)/B_{l=1}(t)|$ with respect to time for a few values of $l$. Note that in the tables only the vacuum multipoles are shown. The multipoles appearing inside the crust are of higher order.

#### *5.2.1. Lowest Eigenmode Case*

Table 3 shows $|B_l(t)/B_{l=1}(t)|_{max}$ for each run. Initially the torroidal current is weaker for this case so high multipoles are expected to be smaller than for the buried case. This can also be seen by comparing tables 3 and 4. As can be expected for the lower initial values of $R_m$ only one or two multipoles are generated but more and more start to build up as $R_m$ gets higher.

#### *5.2.2. Buried case*

Table 4 is similar to table 3 but with the field buried initially. Again the build-up of multipoles increases with $R_m$ but it is also stronger in this case compared to table 3. The lowest order multipoles ($l = 3$ to $\sim 9$) are a substantial fraction of the dipole mode when $R_m = 50, 100$. In particular the $R_m = 100$ case shows the $l = 5,\ 9$ modes to be higher than the 40 % argued by Arons (1993). One must keep in mind the outer parts of the crust are not modelled accurately here and a more detailed analysis could turn out to be more in line with that observational argument. For the $R_m = 100$ case high multipoles $l > 10$ contribute substantially to the magnetic field. However, the contribution from these high order multipoles are only significant within a few tenths of an ohmic time.

## 6. Conclusions

We have investigated the effect of Hall-drift on the magnetic energy evolution of a neutron star crust. We compared the evolution with Halldrift to the one found with only ohmic diffusion and found an effect of up to 30 % on the decay rate. Another interesting finding is that in some cases the torroidal energy acted as an energy reservoir and reinjected energy into the poloidal component after about an ohmic time, resulting in *smaller* field decay with time compared to pure ohmic decay. This is likely to depend on the specific initial conditions and boundary conditions that were used. We also found the vacuum multipole evolution produces some high order multipoles especially when the magnetic Reynold's number is initially strong.

One important limitation in our calculations is that there is always a surface current running within the London depth of the superconductor, a current that does nothing to effect the field energy decay. For future investigations it would therefore be interesting to see the effect of relaxing the inner boundary condition. Instead of having a superconductor there one could investigate the case where the whole spere has some constant density and conductivity. Further, investigating the case of having high angular modes initially and see how significantly the dipole mode would contribute with time should be interesting. In this case the axi-symmetry of the present model could be a severe limitation.

I wish to thank Dong Lai, Andreas Reisenegger, Christopher Pethick, and Peter Goldreich for many interesting discussions and useful suggestions during the course of this project. This work was supported in part by the U. S. National Science Foundation under grants NSF AST93-15133 and AST94-14232, by NASA under grant NAGW-1583, and by the Swedish Natural Science Research Council.

## REFERENCES

Arons, J. 1993, ApJ, 408, 160

Bowers, R.L., Wilson R.J. 1991, Numerical modeling in applied physics and astrophysics, (Boston: Jones-Bartlett)

Chanmugam, G., Sang, Y. 1989, MNRAS, 241, 295

Chanmugam, G. 1992. ARA&A, 30, 143

Goldreich, P., Reisenegger A. 1992, ApJ, 395, 250

Haensel, P., Urpin, V., Yakovlev, D. 1990, A&A, 229, 133

Jackson, J.D. 1975, Classical Electrodynamics, (New York: Wiley)

Jones, P.B. 1988, MNRAS, 233, 875

Lamb. F. 1991, in Frontiers of Stellar Evolution, ed. D. Lambert (San Francisco : Astronomical Society of the Pacific), 299

Landau, L. D., Lifshitz, E.M., Pitaevskii, L.P. 1984, Electrodynamics of Continuous Media, Pergamon Press Oxford

Muslimov A. 1994, MNRAS, 267, 523

Muslimov A., Page D. 1995, ApJ, 440, L77

Naito T., Kojima Y. 1994, MNRAS, 266, 597

Pethick, C. 1992a, in Structure and evolution of neutron stars, eds. D. Pines, R. Tamagaki, S. Tsuruta (Redwood City : Addison-Wesley), 115

Pethick, C. J. 1992b, Rev. Mod. Phys., 64, 1133

Pethick, C. J., Sahrling, M. 1995, ApJ, 453, L29

Phinney, S., Kulkarni, S. 1994, ARA&A, 32, 591

Urpin, V. A., Yakovlev, D. G. 1980, Soviet Ast., 24, 303

Yakovlev, D. G., Shalybkov, D. A.. 1991a, Ap&SS, 176, 171

Yakovlev, D. G., Shalybkov, D. A. 1991b, Ap&SS, 176, 191

Young, E.J., Chanmugam, G., 1995, ApJ, 442, L53

---

This preprint was prepared with the AAS LaTeX macros v4.0.

Fig. 1.— The energy in the crust as a function of time when only the Hall term is kept in equation (3). For initial conditions the buried case, equation (20), was used. The parameter $\tau_{Hall} = t/t_{Hall}$. The calculation was stopped after about 12 $t_{Hall}$ due to features appearing that were too sharp to be resolved. The torroidal energy was then about 27 % of the poloidal energy. A total of 20000 timesteps were performed.

Fig. 2.— The poloidal and torroidal energies as a function of time with the lowest ohmic eigenmode initially and an initial $R_m = 100$. Note the difference to the buried case where the poloidal field energy is comparable to the poloidal field energy.

Fig. 3.— a) The poloidal and torroidal energies as a function of time for the initially buried case with an initial $R_m = 50$. Note how the torroidal energy is pushing energy back into the poloidal component after about half an ohmic time. b) Same as a) but here $R_m = 100$ initially. The torroidal field behaving as a reservoir is here even more apparent than for $R_m = 50$.

Table 1. $E_{Hall}(t)/E_{Ohm}(t)$: Initially Lowest Ohmic Eigenmode
The accuracy, $\epsilon$, for these runs is 0.02 for all cases

| Ohm Time | $R_m = 5$ | 10 | 20 | 50 | 100 |
|---|---|---|---|---|---|
| 0.5 | 1.00 | 1.00 | 0.98 | 0.95 | 0.94 |
| 1 | 1.00 | 0.99 | 0.96 | 0.92 | 0.92 |
| 2 | ··· | ··· | 0.94 | 0.90 | ··· |
| 3 | ··· | ··· | 0.93 | 0.89 | ··· |
| 4 | ··· | ··· | 0.93 | 0.88 | ··· |

Table 2. $E_{Hall}(t)/E_{Ohm}(t)$, Initially Buried Case
The accuracy, $\epsilon$, for these runs ranges from $7\ 10^{-4}$ to $3\ 10^{-3}$

| Ohm Time | $R_m = 5$ | 10 | 20 | 50 | 100 |
|---|---|---|---|---|---|
| 0.5 | 1.00 | 1.01 | 1.00 | 0.95 | 0.96 |
| 1 | 1.00 | 1.00 | 1.00 | 1.03 | 1.33 |
| 2 | 1.00 | 1.00 | 1.00 | 1.18 | ··· |
| 3 | 1.00 | 1.00 | 1.01 | 1.16 | ··· |
| 4 | ··· | 1.00 | 1.01 | 1.15 | ··· |
| 5 | ··· | 1.00 | 1.01 | 1.15 | ··· |
| 10 | ··· | 1.00 | 1.01 | 1.13 | ··· |

Table 3. $|B_l/B_{l=1}|_{max}$: Initially Lowest Ohmic Eigenmode

| $l$ | $R_m = 5$ | 10 | 20 | 50 | 100 |
|---|---|---|---|---|---|
| 3 | $4.9\ 10^{-6}$ | $1.6\ 10^{-1}$ | $4.0\ 10^{-1}$ | $5.1\ 10^{-1}$ | $4.0\ 10^{-1}$ |
| 5 | $1.8\ 10^{-4}$ | $2.0\ 10^{-3}$ | $4.9\ 10^{-2}$ | $5.4\ 10^{-2}$ | $4.9\ 10^{-2}$ |
| 7 | $4.7\ 10^{-6}$ | $1.2\ 10^{-3}$ | $1.4\ 10^{-2}$ | $1.9\ 10^{-2}$ | $1.4\ 10^{-2}$ |
| 9 | $1.7\ 10^{-6}$ | $1.2\ 10^{-4}$ | $2.8\ 10^{-3}$ | $2.8\ 10^{-3}$ | $2.8\ 10^{-3}$ |
| 11 | ... | ... | $4.9\ 10^{-4}$ | $8.0\ 10^{-3}$ | $4.9\ 10^{-4}$ |
| 13 | ... | ... | $6.6\ 10^{-4}$ | $7.3\ 10^{-4}$ | $6.6\ 10^{-4}$ |

Table 4. $|B_l/B_{l=1}|_{max}$: Initially Buried Case

| $l$ | $R_m = 5$ | 10 | 20 | 50 | 100 |
|---|---|---|---|---|---|
| 3 | $5.0\ 10^{-3}$ | $3.7\ 10^{-2}$ | $7.3\ 10^{-2}$ | $1.6\ 10^{-1}$ | $3.3\ 10^{-1}$ |
| 5 | $5.6\ 10^{-5}$ | $3.7\ 10^{-3}$ | $5.2\ 10^{-2}$ | $2.4\ 10^{-1}$ | $7.3\ 10^{-1}$ |
| 7 | $1.4\ 10^{-6}$ | $7.6\ 10^{-4}$ | $1.4\ 10^{-2}$ | $1.4\ 10^{-1}$ | $4.4\ 10^{-1}$ |
| 9 | $3.7\ 10^{-8}$ | $4.1\ 10^{-5}$ | $2.2\ 10^{-3}$ | $9.5\ 10^{-2}$ | $2.5\ 10^{-1}$ |
| 11 | ... | ... | $8.2\ 10^{-4}$ | $4.5\ 10^{-2}$ | $1.5\ 10^{-1}$ |
| 13 | ... | ... | $3.5\ 10^{-4}$ | $2.3\ 10^{-2}$ | $1.0\ 10^{-1}$ |
| 15 | ... | ... | ... | ... | $1.3\ 10^{-1}$ |
| 17 | ... | ... | ... | ... | $1.2\ 10^{-1}$ |
| 19 | ... | ... | ... | ... | $1.1\ 10^{-1}$ |

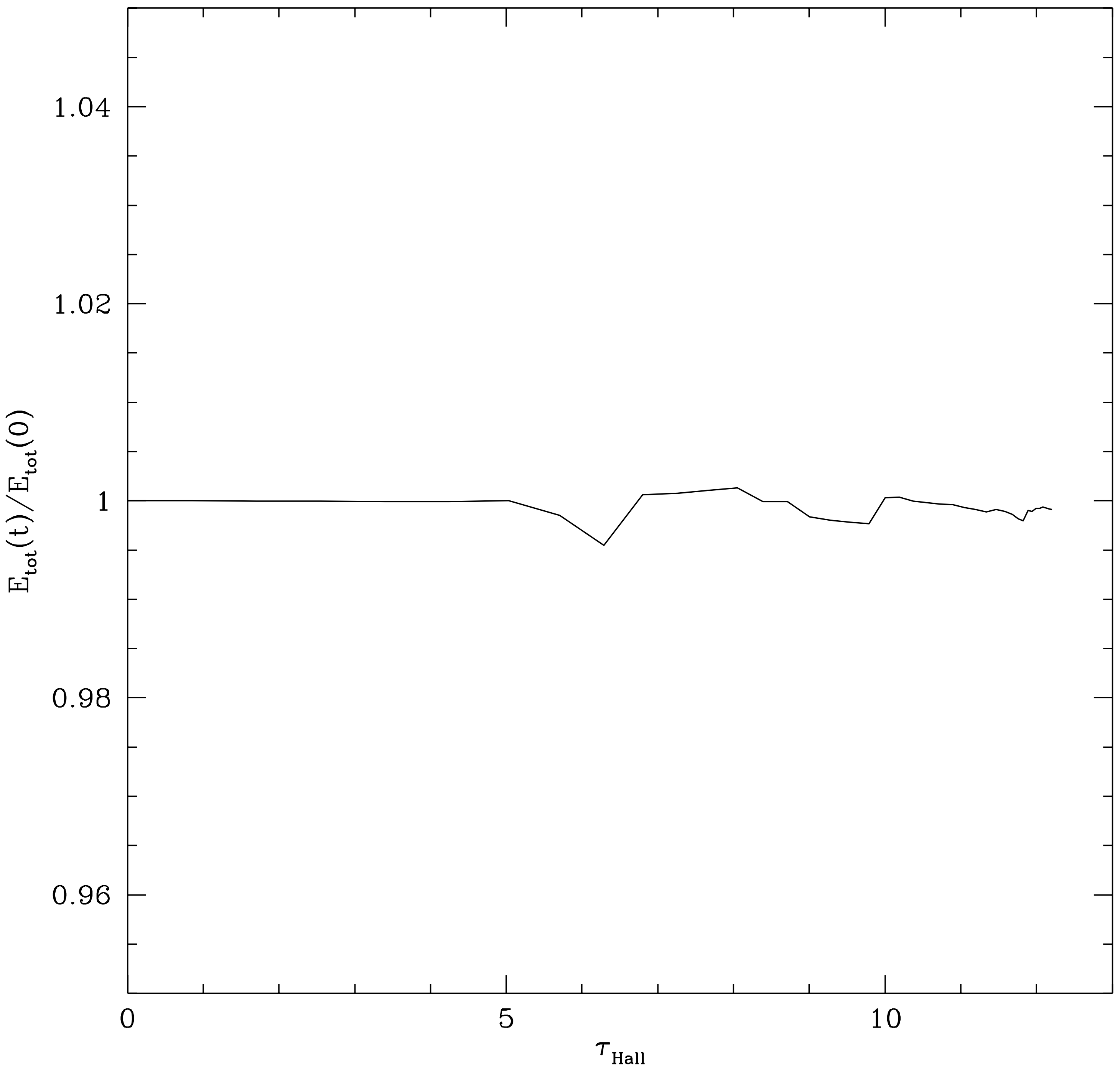

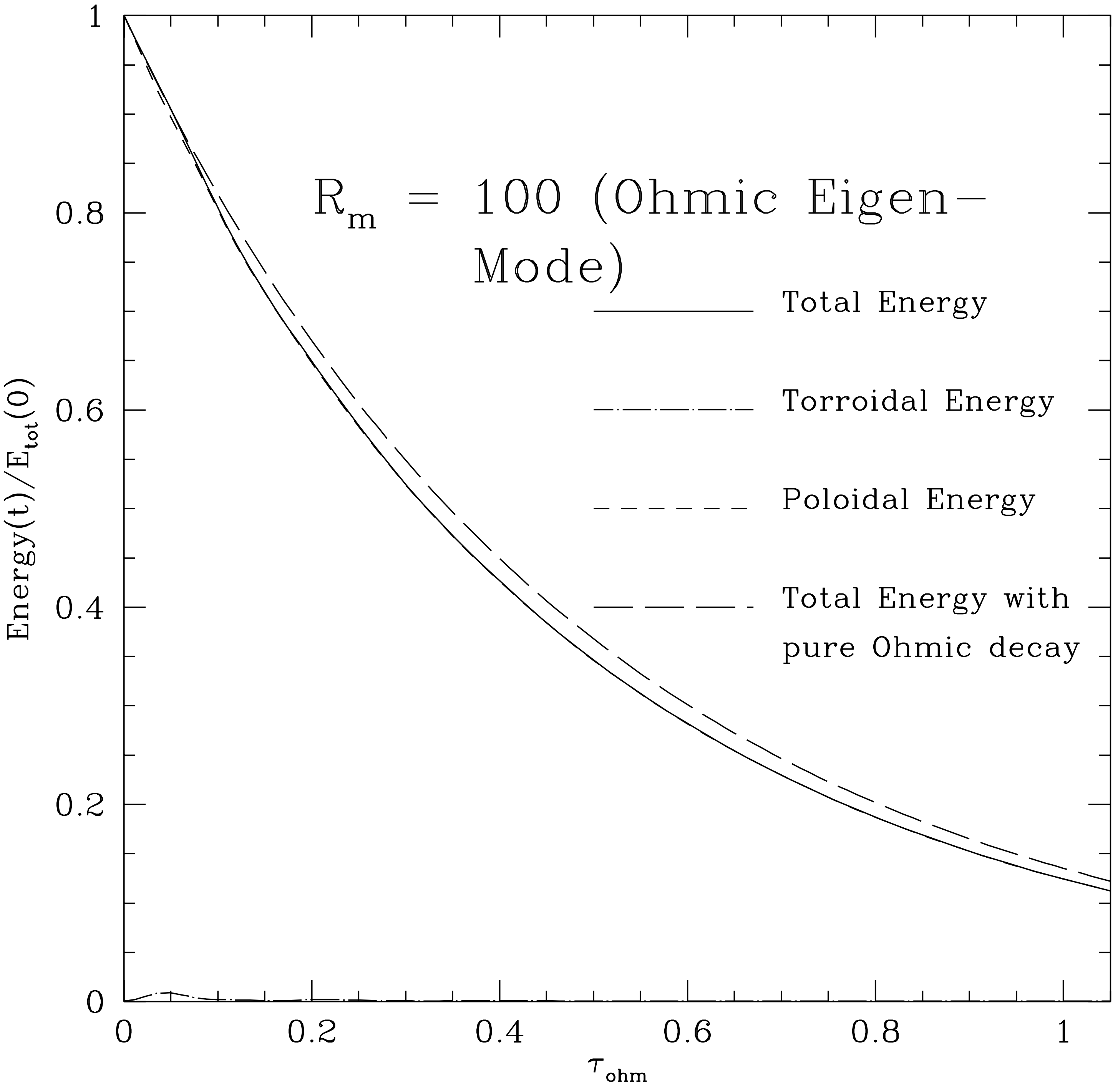

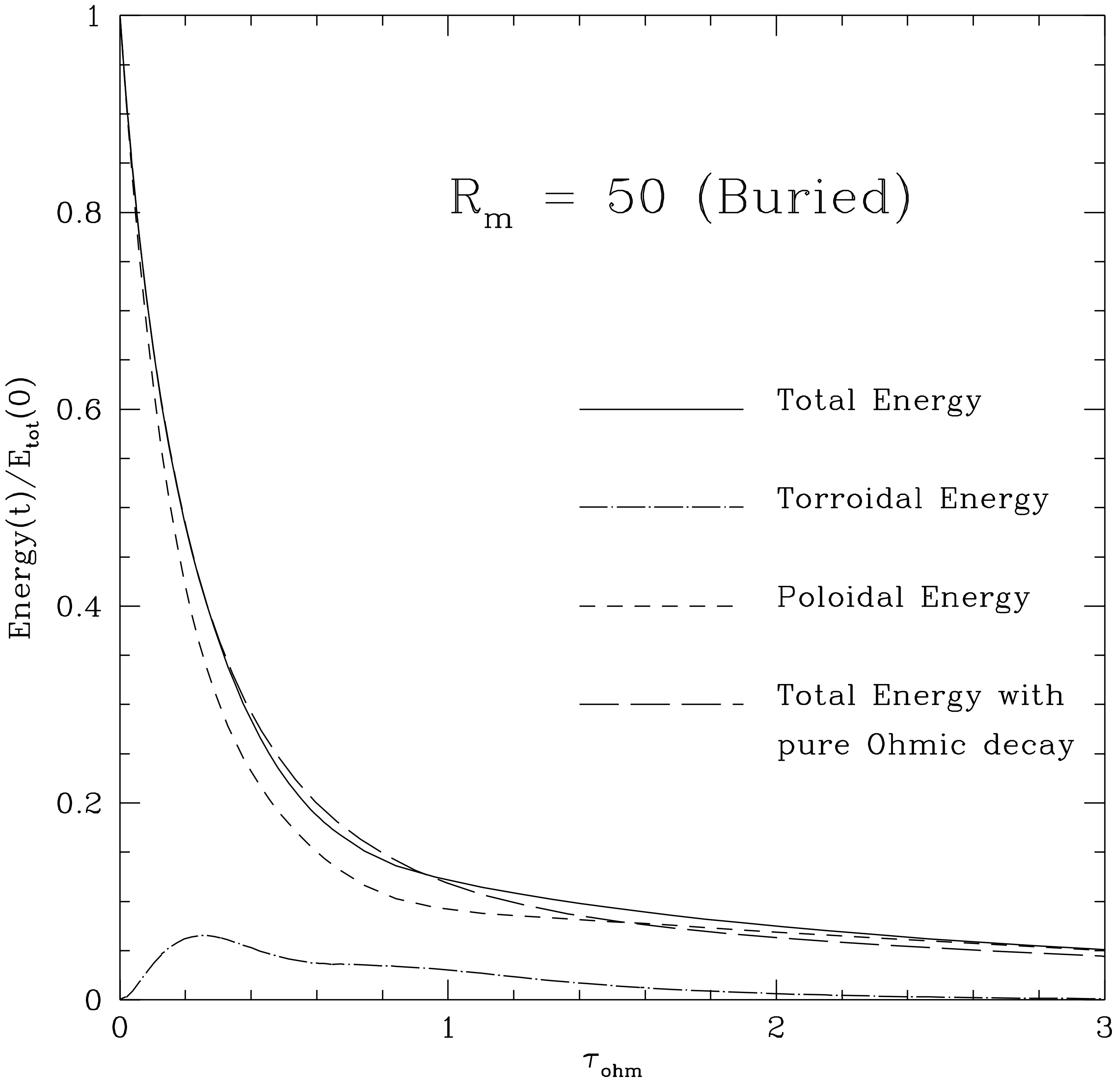

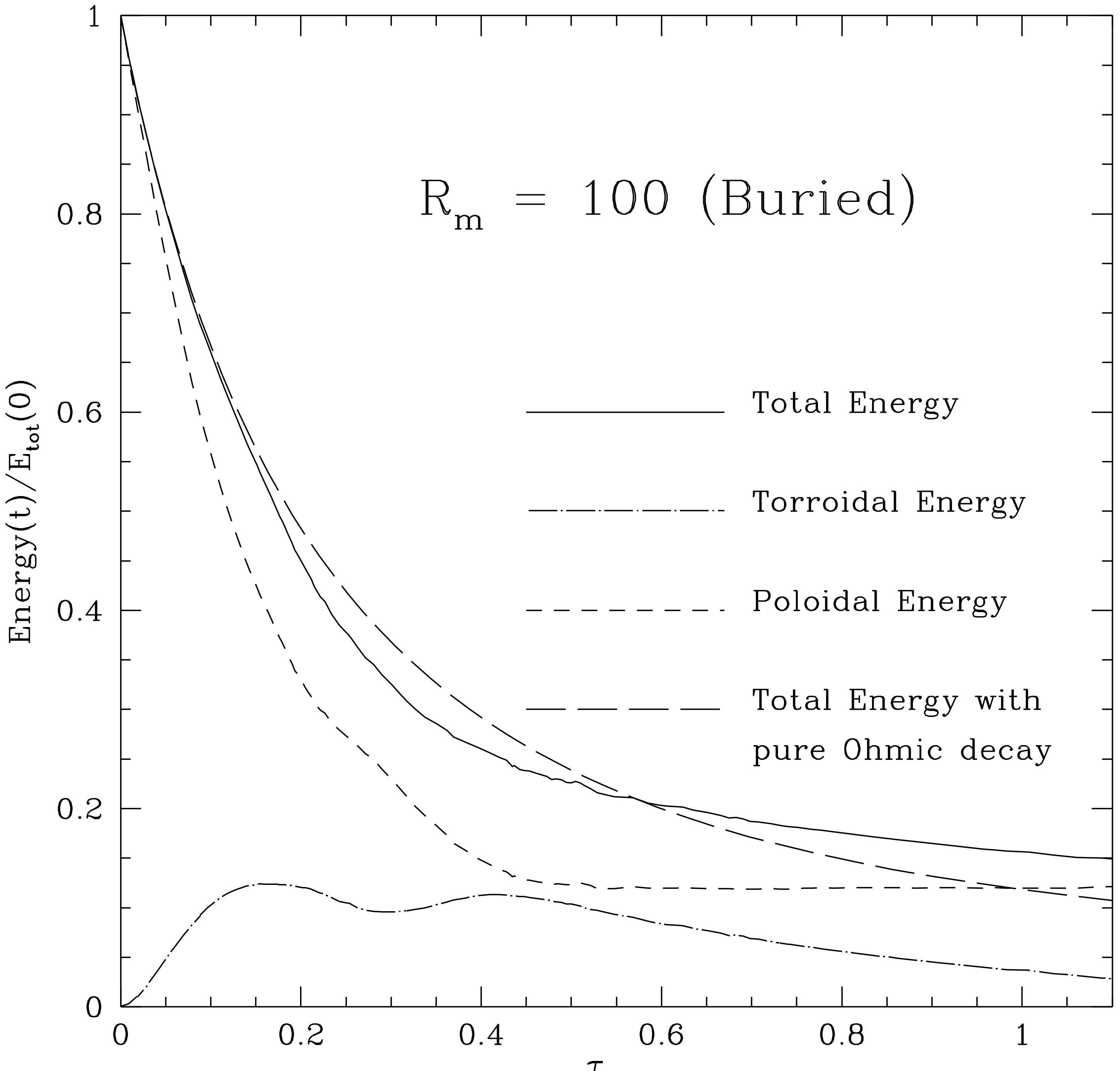